\documentclass[fleqn,twoside]{article}
\usepackage{espcrc2,epsfig}

\usepackage{graphicx}
\usepackage[figuresright]{rotating}

\newcommand{\AmS}{{\protect\the\textfont2
  A\kern-.1667em\lower.5ex\hbox{M}\kern-.125emS}}

\title{RAPID N$_H$ CHANGES IN NGC 4151}
\author{S. Puccetti\address[MCSD]{Universit\'a degli Studi di Roma
    ``Tor Vergata'', Roma, Italy} \address[MCSD]{INAF-Osservatorio Astronomico di Roma, Monteporzio
Catone, Italy}, G.  Risaliti\address[MCSD]{INAF-Osservatorio Astrofisico
di Arcetri, Firenze, Italy}, \address[MCSD]{Harvard-Smithsonian Center for
Astrophysics, Cambridge, USA}, F. Fiore$^a$, M.
Elvis$^d$, F. Nicastro$^d$, G. C.
Perola\address[MCSD]{Universit\'a degli Studi di Roma Tre, Roma,
Italy }, M. Capalbi\address[MCSD]{ASI Science Data Center, c/o
ESA-ESRIN, Frascati, Italy}.}

\begin{document}

\begin{abstract}
We have analyzed two long BeppoSAX observations of the bright Seyfert
galaxy NGC 4151, searching for short timescale ($\sim$10-200 ksec)
X-ray spectral variability. The light curve of a softness ratio,
chosen as most sensitive to pinpoint changes of the column density of
the absorbing gas along the line of sight, shows significant
variations. We try to model these variations by performing a detailed,
time resolved, spectral analysis. We find significant, large (factors of 1.5-6)
variations of the absorber column densities on time scales of 40-200
ksec. These values are 10-100 times shorter than those found by
Risaliti et al. 2002 in a sample of Seyfert 2 galaxies, and provide
strong constraints on the geometry of the obscuring medium.

\vspace{1pc}
\end{abstract}

\maketitle

\section{Introduction}

The geometry of the obscuring matter in Active Galactic Nuclei,
predicted by unified models \cite{ant}, is still unclear. The study of
variability of the absorbing column density (N$_H$) in these objects,
provides invaluable informations on the size and location of the
absorber.  Recently, Risaliti et al. \cite{ris} found N$_H$ variability
on time scales from $\sim 2$ month to $\sim$ 5 years in nearly all the
objects belonging to a sample of bright, Compton-thin Seyfert 2
galaxies with multiple X-ray observations. This variability suggests
that the absorber could be much closer to the nucleus than usually
assumed.

To further investigate this issue, we have analyzed two BeppoSAX long
observations of the brightest Seyfert galaxy, NGC 4151, and present
here some preliminary results.

\hspace{0.cm}
\section{Light curves analysis}

The two BeppoSAX observations were performed on
2001 December, 18 and 1996  July, 6  with the Narrow
Field Instruments: LECS (0.1-10 keV), MECS (1.3-10 keV), and PDS
(13-200 keV).

The total absorbing column density measured in NGC4151 by HEAO1,
EXOSAT, Ginga, ROSAT and ASCA \cite{holt,per,fio,weav,pir,wang,zd}
ranges from a few $10^{22}$ cm$^{-2}$ to a few $10^{23}$ cm$^{-2}$.
Since the corresponding photoelectric cut-off lies in the 2-4 keV
energy range, we have chosen to analyze the light curves of the count
rate in the bands 2-4 keV (MECS), 6-10 keV (MECS) and of their
softness ratio (see Figures 1 and 2), to investigate whether N$_H$
variations might have occurred within each observation. We also
analyzed the light curve of the count rate in the 15-200 keV band
(PDS) to monitor the continuum.  From the analysis of these light
curves we have selected some time intervals where we find indications
that N$_H$ might have undergone significant changes. These intervals
are shown and labelled in figures 1 and 2. To quantify the spectral
variations we have then performed a detailed spectral analysis of
these stretches of data.

\begin{figure}[h]
\includegraphics[width=8cm]{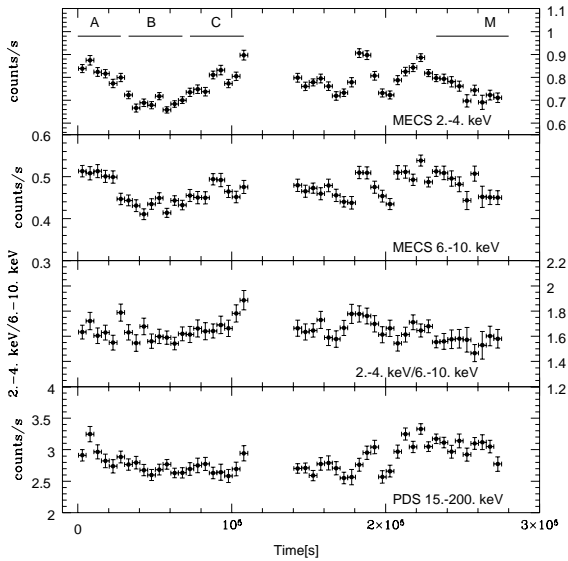}
\caption{Light curves of 2001 December 18 observation. From top to
bottom: MECS 2-4 keV count rate, MECS 6-10 keV count rate, 2-4
keV/6-10 keV softness ratio, PDS 15-200 keV count rate.  The capital
letters at the top of the figure, indicate the time intervals selected
for a time-resolved spectral analysis.}
\end{figure}
\begin{figure}[h]
\includegraphics[width=8cm]{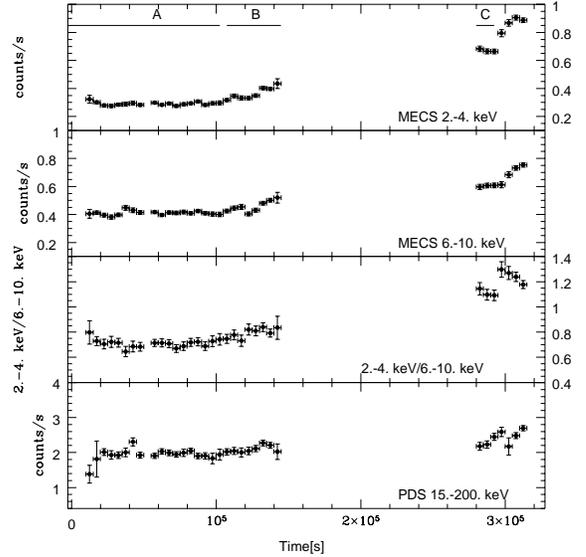}
\caption{Same panels of figure 1 but for the 1996 July 6 observation. 
}
\end{figure}
\section{Spectral analysis}

Standard data reduction was performed using the SAXDAS software
package version 2.0 following Fiore, Guainazzi \& Grandi \cite{fgg}.
Spectral fits were performed using the XSPEC 11.2.0 software package
and public response matrices as from the 1999 December release.  The
PDS data were reduced using the ``variable risetime threshold''
technique to reject particle background \cite{fgg}. LECS and MECS
spectra were rebinned following two criteria:a) to sample the energy
resolution of the detectors with three channels at all energies
whenever possible, and to obtain at least 20 counts per energy
channel. Constant factors have been introduced in the fitting models
in order to take into account the intercalibration systematics between
the instruments \cite{fgg}.  The model adopted for the analysis has
the following components:

\begin{enumerate}
\item A power-law with an exponential high-energy cut-off;

\item a neutral Compton reflection component (modelled with PEXRAV
   \cite{mag});

\item a narrow iron K$_\alpha$ emission line with energy set to 6.38 keV
in the observer frame;

\item two components for the soft excess emission: a thermal
component, and a power law, reflection component.

\item two components for the intrinsic absorber, both assumed neutral:
$N_{H_1}$ which covers the nucleus totally; $N_{H_2}$ which covers the
nucleus only in part, with a covering factor CV
\cite{holt,per,fio,weav,pir,wang,zd}.  The absorber is completed with
the line of sight column density through the Galaxy
(N$_{H_{gal}}$$=$2.17$\times 10^{20}$ cm$^{-2}$\cite{mu}).
\end{enumerate}

Figures 3 and 4 illustrate the results for the two observations.  In
the 2001 December, 18 observation, we find that from A to B to C, both
absorbers and the photon index do not show statistically significative
changes. From C to M the two absorbers N$_{H_1}$ and N$_{H_2}$ change
at a confidence level $\sim 90\%$ by a factor of 2 and 3.3
respectively, while the photon index is constant at a confidence level
$\sim 99 \% $. If the covering factor CV is kept constant throughout
the fit the amplitude of the variations of N$_{H_1}$ and N$_{H_2}$ is
reduced to a factor 1.4 and 2 respectively.  The variations occur on a
timescale of 100-150 ksec.

In the 1996 July, 6 observation, we find that from A to B both
absorbers undergo changes at a confidence level $\sim90\%$, while from
A to C the absorbers change at a confidence level $>99\%$.  From A to
B to C, the photon index is again statistically consistent with a
constant.  The amplitude of the variation of N$_{H_1}$ and N$_{H_2}$
is a factor 6 and 5 respectively.  If the covering factor CV is kept
constant throughout the fit the amplitude of the variations of
N$_{H_1}$ and N$_{H_2}$ is reduced to a factor 4 and 3 respectively. 
The variations occur on
a timescale of $\approx30$ ksec from A to B and $\approx200$ ksec 
from A to C.

\begin{figure}
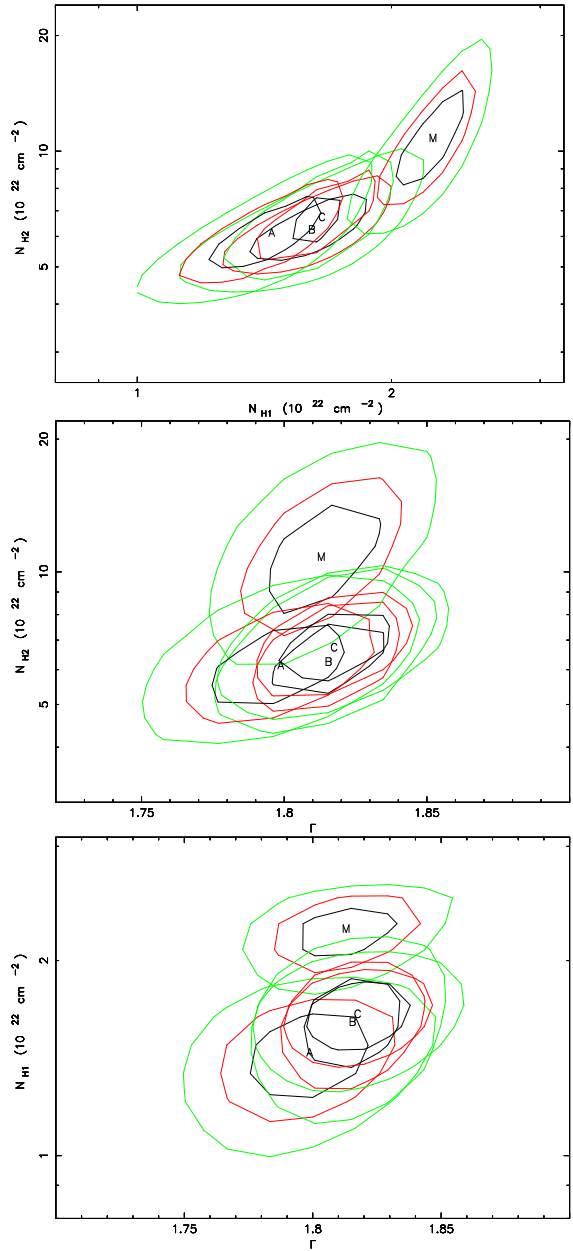

\includegraphics[width=5.5cm,angle=-90]{contour_18dec_tot_nh1_nh2_new.ps}
\includegraphics[width=5.5cm,angle=-90]{contour_18dec_tot_nh2_g_new.ps}
\includegraphics[width=5.5cm,angle=-90]{contour_18dec_tot_nh1_g_new.ps}
\vspace{-0.1cm}
\caption{From top to bottom: 68, 90, 99$\%$ $\chi^2$ confidence
contours for N$_{H_2}$ versus N$_{H_1}$, N$_{H_2}$ versus photon index
$\Gamma$ and N$_{H_1}$ versus $\Gamma$, for the spectra corresponding
to the intervals A, B, C and M in the 2001 December, 18 observation.}
\end{figure}

\begin{figure}
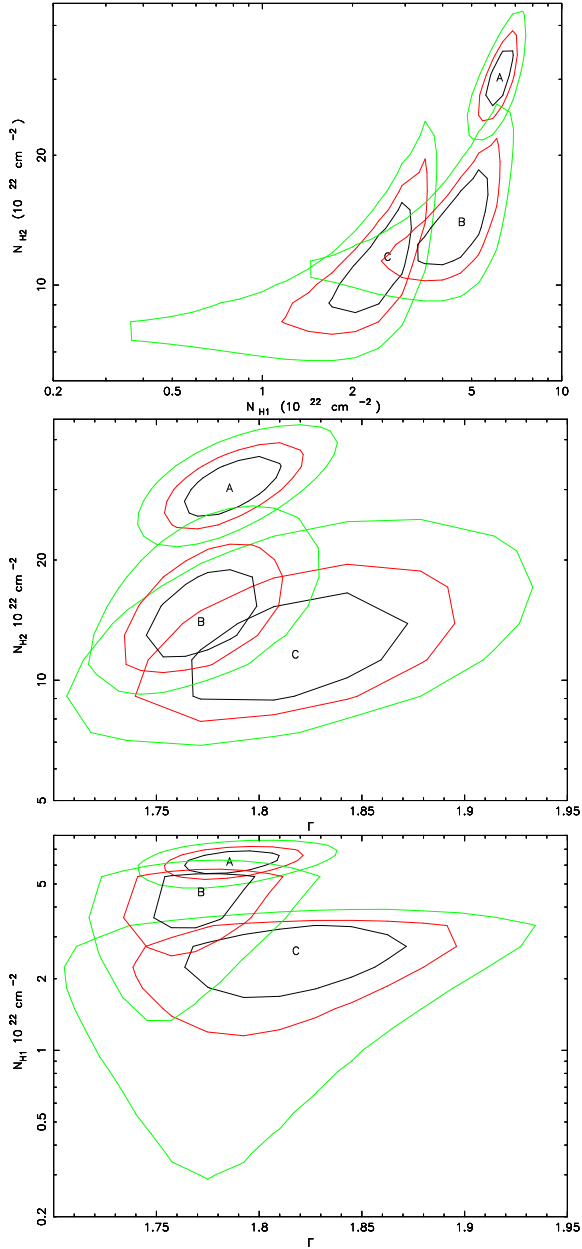

\includegraphics[width=5.5cm,angle=-90]{contour_6jul_tot_nh1_nh2_new2.ps}
\includegraphics[width=5.5cm,angle=-90]{contour_6jul_tot_nh2_g_new.ps}
\includegraphics[width=5.5cm,angle=-90]{contour_6jul_tot_nh1_g_new2.ps}
\caption{From top to bottom: 68, 90, 99$\%$ $\chi^2$ confidence
contours for N$_{H_2}$ versus N$_{H_1}$, N$_{H_2}$ versus photon index
$\Gamma$ and N$_{H_1}$ versus $\Gamma$, for the spectra corresponding
to the intervals A, B and C in the 1996
July, 6, observation.}
\end{figure}

\section{Conclusion}

We have performed a detailed time-resolved spectral study of two long
BeppoSAX observations of the Seyfert galaxy NGC4151.  We find strong
evidences for spectral variability which we model in terms of
variations of two absorbing screens.  The timescales of these
variations are between 40 ksec and 200 ksec, 10-100 times shorter than
the typical timescales found by Risaliti et al.\cite{ris} in a sample
of Compton thin Seyfert 2 galaxies.  If we associate these timescales
to a crossing time, the resulting linear size  
would be of about one light day, which is the size of the innermost part
of the Broad Line Region, as inferred from reverberation mapping in
NGC 4151. This interpretation is supported by the
argument, used by Risaliti et al.\cite{ris}, which assume that the
absorbing medium might be made up of spherical clouds moving with
Keplerian velocities around the central black hole.  This 
requires that the cloud gas density is  of the order of
10$^{9}$ cm$^{-3}$, typical of Broad Line Region clouds \cite{bl}.

\end{document}